\documentclass{elsart}
\usepackage{graphicx}

\begin{document}

\begin{frontmatter}
\title{Determination of the Gamow-Teller Quenching Factor
from Charge Exchange Reactions on ${}^{\bf 90}{\bf Z\lowercase{r}}$}

\author{K.~Yako$^1$,}
\author{H.~Sakai$^{1}$,}
\author{M.B.~Greenfield$^2$,}
\author{K.~Hatanaka$^3$,}
\author{M.~Hatano$^1$,}
\author{J.~Kamiya$^4$,}
\author{H.~Kato$^1$,}
\author{Y.~Kitamura$^3$,}
\author{Y.~Maeda$^1$,}
\author{C.L.~Morris$^5$,}
\author{H.~Okamura$^6$,}
\author{J.~Rapaport$^7$,}
\author{T.~Saito$^1$,}
\author{Y.~Sakemi$^3$,}
\author{K.~Sekiguchi$^8$,}
\author{Y.~Shimizu$^3$,}
\author{K.~Suda$^6$,}
\author{A.~Tamii$^3$,}
\author{N.~Uchigashima$^1$,}
\author{T.~Wakasa$^9$}
\address{
${}^1$Department of Physics, University of Tokyo, Bunkyo, Tokyo 133-0033, Japan
\\
${}^2$International Christian University, Mitaka, Tokyo 181-8585, Japan
\\
${}^3$Research Center for Nuclear Physics, Osaka University,
Ibaraki, Osaka 567-0047, Japan
\\
${}^4$Japan Atomic Energy Research Institute, Tokai, Ibaraki 319-1195, Japan
\\
${}^5$Los Alamos National Laboratory, Los Alamos, NM 87545, USA
\\
${}^6$Department of Physics, Saitama University,
Saitama, Saitama 338-8570, Japan
\\
${}^7$Department of Physics, Ohio University, Athens, Ohio 45701, USA
\\
${}^8$The Institute of Physical and Chemical Research (RIKEN),
Wako, Saitama 351-0198, Japan
\\
${}^9$Department of Physics, Kyushu University, Higashi, Fukuoka
812-8581, Japan
}

\begin{abstract}
        Double differential cross sections between $0^\circ$--$12^\circ$
 were measured for the ${}^{90}{\rm Zr}(n,p)$ reaction at 293~MeV over a
 wide excitation energy range of 0--70~MeV.
        A multipole decomposition technique was applied to the present
 data as well as the previously obtained ${}^{90}{\rm Zr}(p,n)$  data 
 to extract the Gamow-Teller (GT) component from the continuum.
        The GT quenching factor $Q$ was derived by using the obtained
 total GT strengths. 
        The result is $Q=0.88\pm 0.06$, not including an overall
 normalization uncertainty in
 the GT unit cross section of 16\%.
\end{abstract}
\end{frontmatter}
\parskip 1mm
\parindent 6mm
\maketitle

The $(p,n)$ reaction at intermediate energies $(T_p > 100~{\rm MeV})$
provides a highly selective probe of spin-isospin excitations in nuclei
  due to the energy dependence of the isovector part of nucleon-nucleon
  ({\it NN}\/)  $t$-matrices~\cite{FL}.
The $0^\circ$ spectrum of this reaction is marked by the
dominance of the Gamow-Teller (GT) giant resonance (GTGR),
which is 
the $\sigma\tau_\pm$ mode~\cite{Doering,ikeda}.
 There exists a model-independent sum rule,
 $S_{\beta^-}-S_{\beta^+} = 3(N-Z)$, 
 where $S_{\beta^\pm}$ is the total GT strength observed for the
 $\beta^\pm$ type~\cite{ikedasum}.
        Surprisingly, however, only a half of the GT sum rule value was 
 identified from $(p,n)$ measurements in the 1980's on targets
 throughout the periodic table~\cite{Gaarde}.
        This problem, the so-called quenching of the GT strength, has
 been one of the most interesting phenomena in nuclear physics because
 it could be related to non-nucleonic degrees of freedom in
 nuclei.
        At that time it was often assumed that the missing strength 
 was shifted to the
 energy region of the $\Delta$ excitation due to coupling between
 nucleon particle-hole ({\it ph}) and $\Delta$-isobar nucleon-hole
 ($\Delta${\it h}) states~\cite{OsetRho,Ericson}.  
        However, part of the quenching is due to the nuclear
 configuration mixing between 1{\it p}1{\it h} and 2{\it p}2{\it h}
 states~\cite{Shimizu,Bertsch}.  
        To discuss the contribution of 2{\it p}2{\it h} states 
 quantitatively, one should search for
 missing GT strength in the continuum, excitation energy region of
 20--50~MeV, where 
 a significant amount of the 2{\it p}2{\it h} component 
 is predicted~\cite{Shimizu,Bertsch}.

        In 1997, Wakasa {\it et al.}~\cite{WA1} accurately measured the
 ${}^{90}{\rm Zr}(p,n)$ spectra at 295 MeV, the energy at which
 spin-flip cross sections are large, distortion effects are
 minimal~\cite{Shen} and therefore the characteristic shapes of the
 angular distributions for each 
 angular momentum transfer ($\Delta L$) are most distinct. 
        They successfully identified the GT strength in the
 continuum region through multipole decomposition (MD) analysis which
 extracted the $\Delta L =0$ component from the cross sections~\cite{WA1}.
        They obtained a GT quenching factor,
        defined as 
 \(\displaystyle Q\equiv\frac{S_{\beta^-}-S_{\beta^+}}{3(N-Z)}\),
 of $0.90\pm0.05$, where
 the error is due to the uncertainty of the MD analysis~\cite{WA1}.
	As discussed in Ref.~\cite{WA1}, the main source of the
 systematic uncertainties is the overall normalization, i.e., the 
 GT unit cross section and the $S_{\beta^+}$ value.
        The uncertainty of the GT unit cross section amounts to 16\%, but
 will be reduced to $\sim5\%$ by the ongoing systematic analysis of new
 $\hat{\sigma}_{\rm GT}$ data at 295 MeV~\cite{Sasano}. 
         On the other hand, the uncertainty in the $S_{\beta^+}$ value
 is difficult to properly assess.
        A $S_{\beta^+}$ value of $1.0\pm0.3$ used in Ref.~\cite{WA1}
 was obtained 
 from a similar, but simpler MD analysis of ${}^{90}{\rm Zr}(n,p)$
 data at 198 MeV by Raywood {\it et al.}~\cite{Raywood}.
        Raywood {\it et al.}\ found a significant amount of the monopole
 ($\Delta L=0$) cross sections in the continuum in the region of
 $E_x=8\sim31$~MeV, which corresponds to a GT strength of $\sim$5, but
 attributed all to the isovector spin-monopole (IVSM)
 strength~\cite{Raywood}. 
        Thus, the quenching factor was subject to uncertainties of both
 the $\beta^+$ strength in the continuum and contributions of the IVSM
 component. 
        To reduce those systematic uncertainties, it is essential to
 have accurate $(n,p)$ data at the same energy as the $(p,n)$ data, and
 to perform consistent analyses on both sets of data.
        In this letter the measurement of the
 ${}^{90}{\rm Zr}(n,p)^{90}{\rm Y}$ reaction at 293~MeV is reported.
        Performing a consistent analysis on both the $(p,n)$ and $(n,p)$
 data, we have derived a reliable GT quenching factor.

\begin{figure}
\includegraphics[width=7.0cm]{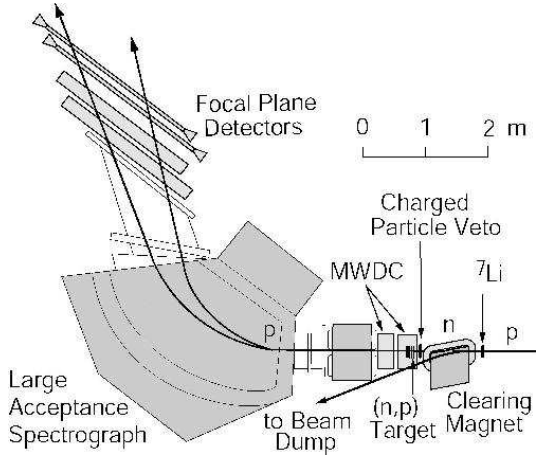}
\caption{
        A schematic view of the $(n,p)$ facility at RCNP.
\label{fig_fac}
}
\end{figure}

        The measurement was performed with the $(n,p)$ facility~\cite{Yako}
 at the Research Center for Nuclear Physics (RCNP).
        A schematic view of the $(n,p)$ facility is shown in
 Fig.~\ref{fig_fac}. 
        A nearly mono-energetic neutron beam was produced by the
 ${}^7{\rm Li}(p,n)$ reaction at 295 MeV. 
        The primary proton beam, after going through the $^7{\rm Li}$
 target, was deflected away by the clearing magnet to a Faraday cup in
 the floor.
        The typical beam intensity was 450~nA and the thickness
 of the ${}^7{\rm Li}$ target was 320 mg/cm$^2$.
        About $2\times 10^6/{\rm s}$ neutrons bombarded the target
 area of $30^W\times 20^H~{\rm mm^2}$ located 95 cm downstream from the
 $^7{\rm Li}$ target. 
        Three ${}^{90}{\rm Zr}$ targets with thicknesses of
 485, 233, and 215~${\rm mg/cm^2}$ and 
 a polyethylene (${\rm CH}_2$) target with 
 a thickness of 46 ${\rm mg/cm^2}$ were mounted in a multiwire drift
 chamber (MWDC). 
        Wire planes placed between the targets detected outgoing protons
 and enabled us to determine the target in which the reaction had occurred.
        Charged particles coming from the beam line were rejected by a
 veto scintillator with a thickness of 1~mm. 
        The scattering angle of the $(n,p)$ reaction was determined by
 the information from the target MWDC and another 
 MWDC installed at the entrance of the Large Acceptance Spectrometer
 (LAS)~\cite{MatsuokaNoro}.
        The outgoing protons were momentum-analyzed by LAS and were
 detected by the focal plane detectors~\cite{NoroOkihana}.
        The number of ${}^1{\rm H}(n,p)$ events from the ${\rm CH_2}$
 target was compared to the {\small SAID}~\cite{SAID} calculated
 ${}^1{\rm H}(n,p)$ cross sections for normalization of the neutron beam
 flux. 
        Blank target data were also taken for background subtraction.

\begin{figure}
\includegraphics[width=8.0cm]{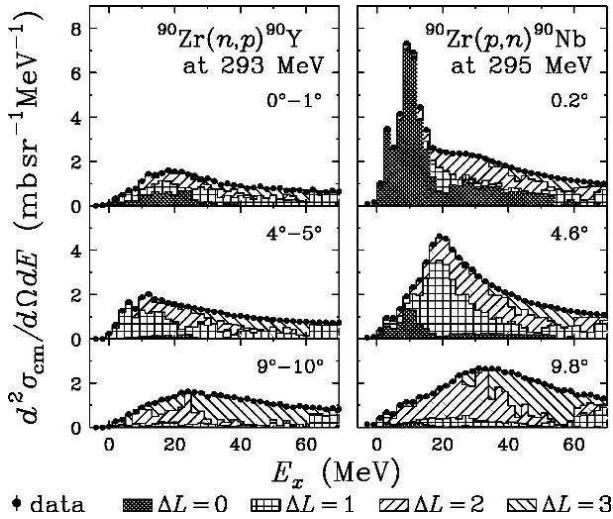}
\caption{
        Double differential cross sections for the ${}^{90}{\rm Zr}(n,p)$ 
 (left panel)~\cite{WA1} and ${}^{90}{\rm Zr}(p,n)$ (right panel) 
 reactions.
        The histograms show the results of the MD analyses.
\label{fig_raw}
}
\end{figure}

        We have obtained double differential cross sections up to 70 MeV
 excitation energy over an angular range of $0^\circ$--$12^\circ$ in
 the laboratory frame.
        The data have been analyzed in 1-degree bins.
        The ${}^{90}{\rm Zr}(n,p)$ spectra at three of the twelve 
 angles are shown
 in the left panel in Fig.~\ref{fig_raw} by the solid dots.
        The overall energy resolution is 1.5~MeV, mainly originating
 from the target thicknesses and the energy spread of the beam.
        The angular resolution is 10~mr which is dominated by multiple
 scattering effects in the ${}^{90}{\rm Zr}$ targets.
        In addition to the statistical uncertainty of $\sim 2$\% per 
 2-MeV excitation energy bin, there is a systematic uncertainty of 5\%,
 where the main contributions are that of target thicknesses (4\%) and 
 the angular distribution of the $n+p$ cross section taken from the 
 phase-shift analysis (2\%).
        The right panel in Fig.~\ref{fig_raw} shows the ${}^{90}{\rm
 Zr}(p,n)$ spectra~\cite{WA1}. 
        At $0^\circ$ the $(p,n)$ cross sections are larger than the
 $(n,p)$ cross sections not only in the GT resonance region but also in the 
 high excitation energy region of $E_x=70~{\rm MeV}$ due to excess 
 neutrons in ${}^{90}{\rm Zr}$.

        The MD analyses were performed on the $(p,n)$~\cite{WA1}
 and $(n,p)$ excitation energy spectra to obtain GT strengths.
        Details of the MD analysis are given in Ref.~\cite{WA1}.
        For each excitation energy bin between 0~MeV and 70~MeV, the
 experimentally obtained angular distribution $\sigma^{\rm
 exp}(\theta_{\rm cm},E_x)$ has been fitted using the least-squares
 method with the linear combination of calculated distributions,
\begin{equation}
\sigma^{\rm calc}(\theta_{\rm cm},E_x)=\sum_{J^\pi}a_{J^\pi}\sigma^{\rm calc}_{{ph};J^\pi}(\theta_{\rm cm},E_x),
\label{eq_MDA0}
\end{equation}
 where the variables $a_{J^\pi}$ are fitting coefficients all of which 
 have positive values.
        The calculated angular distributions for each spin and parity transfer
 $\sigma^{\rm calc}_{{ph};J^\pi}(\theta_{\rm cm},E_x)$ have been obtained
 using the distorted wave impulse approximation (DWIA) calculations
 described below. 

        The DWIA calculations were performed 
 with the computer code {\small DW81}~\cite{dw81} for the following
 $J^\pi$ transfers:
 $1^+(\Delta L=0)$, $0^-, 1^-, 2^-(\Delta L=1)$, $3^+(\Delta
 L=2)$, and $4^-(\Delta L=3)$.
        The one-body transition densities were calculated from pure
 1$p$1$h$ configurations. 
         The $(1g_{7/2},1g_{9/2}^{-1})$ and $(1g_{9/2},1g_{9/2}^{-1})$
 configurations were used to calculate the GT transitions in the 
 analyses of both the $(p,n)$ and the $(n,p)$ spectra.
        For the transitions with $\Delta L \geq 1$ in the $(p,n)$ channel,
 the active proton particles were restricted to the $1g_{9/2}$,
 $1g_{7/2}$, $2d_{5/2}$, $2d_{3/2}$, $1h_{11/2}$, or $3s_{1/2}$ shells,
 while the active neutron  holes were restricted to the $1g_{9/2}$,
 $2p_{1/2}$, $2p_{3/2}$,  $1f_{5/2}$, or $1f_{7/2}$ shells by assuming ${}^{40}{\rm Ca}$ to be the core.
        In the analysis of the $(n,p)$ spectra,
        the active neutron
 particles were restricted to the $1g_{7/2}$, $2d_{5/2}$,
 $2d_{3/2}$, $1h_{11/2}$, or $3s_{1/2}$ shells, while the active proton
 holes to $2p_{1/2}$, $2p_{3/2}$, $1f_{5/2}$, or $1f_{7/2}$.
        The optical model potential (OMP) parameters for proton
 were taken from Ref.~\cite{CooperHama}.
        The OMP parameters for neutron were also taken from
 Ref.~\cite{CooperHama}, but without the Coulomb term.
        The effective {\it NN} interaction was taken from the $t$-matrix
 parameterization of the free {\it NN} interaction by Franey and Love at
 325 MeV~\cite{FL}.
        It should be noted that DWIA calculations using this
 parameter set better reproduce the polarization
 transfer $D_{NN}(0^\circ)$ for the ${}^{90}{\rm Zr}(p,n)$ reaction 
 than those at 270~MeV~\cite{WA1}.
        The radial wave functions were generated from a Woods-Saxon (WS)
 potential~\cite{WoodsSaxon}, adjusting the depth of central potential $V_0$ 
 to reproduce the binding energies~\cite{Kasagi,Horton,Preedom,Graue}.
        The unbound particle states
 were assumed to have a shallow binding energy to simplify the
 calculations. 
        For a given $J^\pi$ transfer, the shapes of the angular 
 distributions depend on the 1{\it p}1{\it h} configurations. Thus, all
 combinations of 1{\it p}1{\it h} configurations were examined by the 
 $\chi^2$-minimization program and the optimal combination was obtained
 for each excitation energy bin.

        Results of the MD analyses are shown in Fig.~\ref{fig_raw}.
        The obtained $\Delta L=0$ component in the $(p,n)$ spectra 
 has a large contribution not only in the GTGR
 region, but also in the high excitation energy region up
 to 50~MeV~\cite{WA1}. 
        The $\Delta L=0$ component  of the cross section, $\sigma_{\Delta 
 L=0}(q,\omega)$, is proportional to the GT strength $B({\rm
 GT})$~\cite{Gaarde} such that, 
\begin{equation}
\sigma_{\Delta L=0}(q,\omega) = \hat{\sigma}_{\rm GT} F(q,\omega) B({\rm GT}),
\label{eq_lin}
\end{equation}
 where $\hat{\sigma}_{\rm GT}$ is the GT unit cross section~\cite{WA1} 
 and $F(q,\omega)$ is the kinematical correction factor~\cite{TaddeucciDWIA}. 
        The GT unit cross section has been determined so that 
 the $S_{\beta^-}$ value up to the GTGR region 
 of $E_x<16~{\rm MeV}$ becomes $18.3\pm3.0$~\cite{Gaarde} as 
 described in Ref.~\cite{WA1}. 
        The obtained value is 
 $\hat{\sigma}_{\rm GT}=3.5\pm0.6~{\rm mb/sr}$, which is consistent
 with $3.6\pm0.6$~mb/sr, the value used in Ref.~\cite{WA1}.

\begin{figure}
\includegraphics[width=8cm]{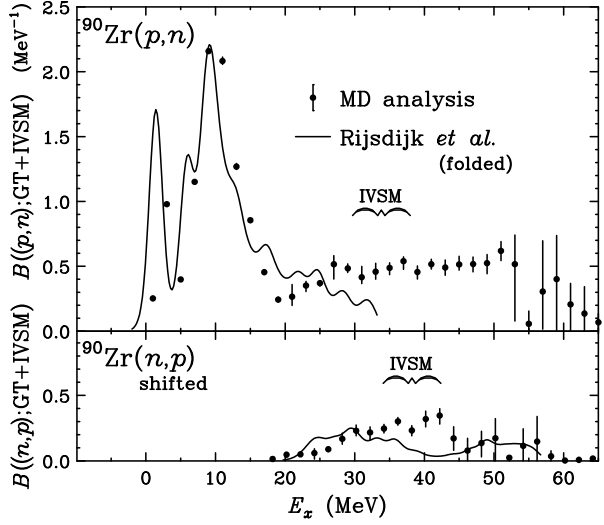}
\caption{\label{fig_bgt}
        GT plus IVSM strength
distributions obtained by the MD analysis of the
 ${}^{90}{\rm Zr}(p,n)$ and ${}^{90}{\rm Zr}(n,p)$ reactions (in GT unit).
  The ${}^{90}{\rm Zr}(n,p)$ spectrum is shifted by $+18$~MeV.
  The curves are taken from Ref.~\cite{Rijsdijk}. 
  The energy regions of IVSM excitation are indicated by braces. 
See text for details.
}
\end{figure}

       The strength distributions
 are shown in Fig.~\ref{fig_bgt}.
       Here the contribution from the isobaric analogue state (IAS) at
 5.1~MeV, corresponding to $0.7\pm0.1$ in GT unit~\cite{WA1}, is
 already subtracted. 
        The strength is denoted as $B({\rm GT+IVSM})$ 
 because it contains the IVSM component~\cite{Raywood}.
        The error bars are the $\pm1\sigma$ confidence limits 
 obtained by a Monte-Carlo simulation,
 where the $\chi^2$ minimization is performed for synthetic data sets
 generated by replacing the actual data set in accordance with the
 statistical errors~\cite{bootstrap}. 
        The MD analysis of the $(p,n)$ spectra becomes unstable above 
 50~MeV excitation~\cite{WA1} and reanalysis with larger bin width has not 
 improved the stability.
        Therefore we set the upper limit of the excitation energy to 
 50~MeV.
        The integrated strength thus obtained is $S_{(p,n);{\rm
 GT+IVSM}}=33.5 \pm 0.6({\rm stat.}) \pm 0.4({\rm MD})\pm 4.7
 (\hat{\sigma}_{\rm GT})$ up to 50~MeV excitation energy, neglecting the
 uncertainty in the subtraction of the IAS contribution.
        The first error is the statistical error of the MD analysis.
        The systematic uncertainty due to the input parameters for DWIA
 calculations has been evaluated by using the wave functions generated from a
 potential by the relativistic Hartree approach~\cite{Horowitz} or 
 with other OMPs~\cite{Shen,Horowitz,RH}, and it is estimated to be $\pm0.4$.
        The third error reflects the error in the GT unit cross
 section. 
        The distribution of $\beta^+$ strength in
 Fig.~\ref{fig_bgt} is  
 shifted by $+18$~MeV, 
 accounting for the Coulomb displacement
 energy and the nuclear mass difference.
        The strength integrated up to 32~MeV excitation of ${}^{90}{\rm
 Y}$, or $E_x=50$~MeV in Fig.~\ref{fig_bgt}, is  
 $S_{(n,p);{\rm
 GT+IVSM}}=5.4\pm0.4({\rm stat.})\pm 0.3({\rm MD})\pm0.9(\hat{\sigma}_{\rm GT})$.  

	The curves in Fig.~\ref{fig_bgt} show the 
 theoretical predictions of GT strength distribution by employing the
 dressed particle random phase
 approximation (DRPA) model~\cite{Rijsdijk}, 
 folded by a Gaussian distribution to simulate
 the energy resolution of the measurement.
        The agreement between the experiment and the theory is excellent, 
 except in the excitation energy region around 30--40 MeV, where the IVSM
 resonance is important~\cite{Raywood,HamamotoSagawa}. 
 The IVSM resonance 
 is the $2\hbar\omega$ excitation via the $r^2\sigma\tau_\pm$ operator 
 and has the same spin and parity
 transfer as the GT resonance.
        Since the angular distribution of the IVSM transition has a
 forward peaking shape similar to that of the GT
 transition~\cite{Auerbach,Prout}, the present 
 MD analysis cannot discriminate these two components. 

         Hamamoto and Sagawa studied the IVSM modes 
 for both the ${}^{90}{\rm Zr}(p,n)$ and ${}^{90}{\rm Zr}(n,p)$ 
 reactions~\cite{HamamotoSagawa}. 
         They pointed out that the response function
 to the operator $r^2\sigma\tau_\pm$ calculated for the $2\hbar\omega$
 excitation contains a small GT component because
 of the difference of the neutron and proton one-particle wave
 functions with the same quantum numbers.
         This GT component has to be subtracted to obtain the pure IVSM
 strength~\cite{HamamotoSagawa}.  
         Although interference exists between the GT and 
 the IVSM modes, the cross section associated with the pure IVSM 
 component is estimated in this work and its contribution is subtracted
 incoherently from each spectrum since the distribution of the GT
 strength in the IVSM resonance region is unknown. 
        The DWIA calculations have been performed by assuming that the
 strengths are fully exhausted in the state with central 
 energies reported to be 35~MeV in the $(p,n)$ spectrum~\cite{Prout}, 
 corresponding to 19~MeV excitation of ${}^{90}{\rm Y}$, and
 $E_x=37$~MeV in Fig.~\ref{fig_bgt}, in the $(n,p)$ spectrum~\cite{HamamotoSagawa}. 
        The transition densities are obtained by the procedure of Cond\'{e}
 {\it et al.}~\cite{Conde}.
        The small GT components are explicitly eliminated by modifying the
 radial wave functions of the final states. 
        The obtained IVSM strengths are within 10\% of the theoretical
 prediction~\cite{HamamotoSagawa}. 
        The calculated IVSM cross sections at $0^\circ$ for the $(p,n)$ and 
 the $(n,p)$ channels are $6.9\pm1.5$~mb/sr ($4.2\pm0.9$ GT units with
 $F(q,\omega)=0.47$) and $5.3\pm0.6$~mb/sr ($2.5\pm 0.3$ GT units with $F(q,\omega)=0.61$), respectively.
        The uncertainties are mainly due to the choice of OMP parameters. 
        By subtracting these values from $S_{\rm
 GT+IVSM}$, the actual GT strengths of 
        $S_{\beta^-}=29.3 \pm 0.5({\rm
 stat.}) \pm 0.4({\rm MD})\pm 0.9({\rm IVSM}) \pm 4.7 (\hat{\sigma}_{\rm GT})$
 and $S_{\beta^+}=2.9 \pm 0.4({\rm stat.}) \pm 0.3({\rm MD})\pm 0.3({\rm
 IVSM}) \pm 0.5 (\hat{\sigma}_{\rm GT})$ have been obtained.
        This new $S_{\beta^-}$ value is consistent with but slightly
 higher than the one reported in Ref.~\cite{WA1}. 
        The $S_{\beta^+}$ value agrees well with the DRPA prediction of 
 $S_{\beta^+}=3.2$~\cite{Rijsdijk}. 

        The present $\beta^+$ strength may be compared with previously
 reported results that employed similar MD techniques on the $^{90}{\rm
 Zr}(n,p)$ spectra. 
        Cond\'{e} {\it et al.}~\cite{Conde} obtained $S_{\beta^+}=1.7\pm
 0.2$ up to 10~MeV excitation at $T_n=98$~MeV while Raywood {\it et
 al.}~\cite{Raywood} obtained $1.0\pm 0.3$ up to $\sim 8~{\rm MeV}$ at
 $T_n=198$~MeV. 
        The errors given above do not include the uncertainty in the GT
 unit cross section. 
        The $\beta^+$ strengths obtained in this work up to 8~MeV and 
 10~MeV 
 excitation are $S_{\beta^+}=0.4\pm0.1({\rm stat.})\pm0.1(\hat{\sigma}_{\rm
 GT})$ and $0.7\pm0.1({\rm stat.})\pm0.1(\hat{\sigma}_{\rm
 GT})$, respectively. 
 These $S_{\beta^+}$ values are significantly smaller than those obtained by 
 Raywood {\it et al.}\ or by Cond\'{e} {\it et al.}
       We note that the analysis of the $T_n=98$~MeV data may suffer
 from ambiguities in the reaction mechanism due to multi-step
 processes while the MD analysis at 200~MeV may suffer from the
 ambiguity due to the distortion effects which 
 are larger than those at 300 MeV.

        The consistent analyses of both (p,n) and (n,p) spectra 
 yield a quenching factor of $Q = 0.88 \pm 0.02({\rm stat.}) \pm
 0.05({\rm syst.}) \pm 0.01({\rm MD}) \pm 0.02({\rm IVSM})$, 
 where the systematic uncertainty of the normalization in 
 the cross section (5\%) is also indicated.
        An uncertainty in the GT unit cross section of 16\% is not
 included. 
        It should be noted that since the errors are correlated, the
 combined systematic errors are smaller than their geometric sum. 
        Thus quenching of GT strength up to 50~MeV due to 
 coupling between {\it ph} and $\Delta h$ states
 becomes significantly smaller when 2{\it p}2{\it h}
 contributions are properly accounted for~\cite{Bertsch,WA1}. 

\begin{figure}
\includegraphics[width=5.6cm]{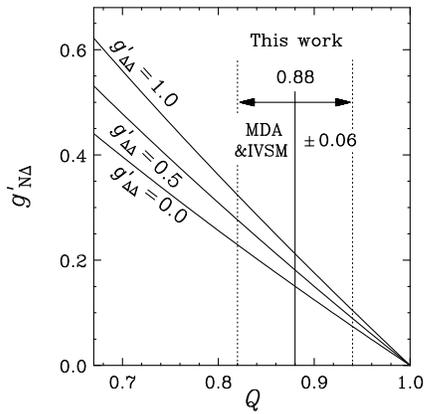}
\caption{\label{fig_gndelta}
       LM parameter $g'_{\rm N\Delta}$ in Chew-Low model estimated as a
 function of quenching factor $Q$. 
 The finite size effect of ${}^{90}{\rm Zr}$ nucleus is not taken into account.
}
\end{figure}

         The interpretation of this small quenching
 in terms of the short range correlation in nuclei is particularly
 interesting.
         Assuming that the missing GT strength of $\sim 10$\% is attributed 
 to the $\Delta$ excitation, one can derive
 the Landau-Migdal (LM) parameter $g'_{\rm N\Delta}$ which describes
 the short range correlations  
 for {\it ph} $\rightarrow \Delta${\it h} transitions
 in the $\pi+\rho+g'$ model~\cite{SuzukiSakai}.
        The $g'_{\rm N\Delta}$ value deduced by using an RPA model in
 {\it ph} and $\Delta${\it h} spaces~\cite{SuzukiSakai} with the 
 Chew-Low~\cite{ChewLow} coupling constant ($f_\Delta/f_\pi=2$) is shown in
 Fig.~\ref{fig_gndelta} as a function of $Q$.
	If we take here $Q=0.88\pm0.06$, combining the
 uncertainty of MD analysis and that of IVSM contribution in quadrature,
        then we derive $g'_{\rm N\Delta}=0.18\pm0.09$ assuming $g'_{\rm
 \Delta\Delta}=0.5$~\cite{Dickhoff}. 
        Arima {\it et al.}\ have examined the finite size
 effects of the ${}^{90}{\rm Zr}$ nucleus by taking the finite
 range interaction due to $\pi$- or $\rho$-exchange into account~\cite{Arima2}. 
        If their argument is employed here, the $g'_{\rm N\Delta}$ value
 increases by 0.07 for the same $Q$. 
        Therefore it is reasonable to assume the $g'_{\rm N\Delta}$
 value to be $0.25\pm0.09$, which is significantly smaller than that
 assuming the universality relation of the LM parameters, i.e., 
$g'_{\rm NN}=g'_{\rm
 N\Delta}=g'_{\Delta\Delta}=0.6$--0.8~\cite{OsetTokiWeise}.  
        This result is consistent with those obtained using
 the coupled channel $G$-matrix calculations~\cite{Dickhoff,SagawaLeeOhta}.
        An important consequence of such a 
 small $g'_{\rm N\Delta}$ value is the enhancement of the pion
 correlation in nuclei.
        According to the RPA prediction by Tatsumi {\it et
 al.}~\cite{Nakano} the critical density of the pion condensation
 becomes $(1.9\pm0.3)\rho_0$, just a half of that 
 predicted by employing the universality assumption.

        In summary, we measured the double differential cross
 sections between $0^\circ$--$12^\circ$ for the ${}^{90}{\rm Zr}(n,p)$
 reaction at 293~MeV in an excitation energy region of 0--70~MeV to
 study the $\beta^+$ GT strengths in the continuum. 
        The MD technique was applied to the measured cross sections to
 extract the $\Delta L=0$ cross section in the continuum.
        After subtracting the IVSM contribution a total GT
 strength of $S_{\beta^+}=2.9 \pm 0.4 ({\rm stat.}) \pm 0.3 ({\rm
 syst.}) \pm0.3({\rm MD}) \pm 0.3({\rm IVSM}) \pm
 0.5 (\hat{\sigma}_{\rm GT})$ up to 32~MeV excitation was obtained. 
        A revised and consistent analysis of the $(p,n)$~\cite{WA1} and $(n,p)$
 reaction data from ${}^{90}{\rm Zr}$ yield a reliable quenching factor of
 $Q=0.88\pm0.06$, not including the uncertainty of the GT unit cross
 section of 16\%.
        The error includes the uncertainty of the estimated IVSM
 contribution as well as error in the data and the MD analysis. 
        This work is the first attempt to deduce the GT quenching factor 
 by accurately taking into account GT strengths in the continuum
 region. 

\ack
     We wish to acknowledge the outstanding support of the accelerator group
 of RCNP. 
     We thank I.~Hamamoto, H.~Sagawa and T.~Suzuki for valuable discussions.
     This work was supported financially in part by the Grant-in-Aid for
 Scientific Research No.~10304018 of Ministry of Education, Science,
 Culture and Sports of Japan.
\bibliographystyle{elsart-num}
\bibliography{paper_yako}

\begin{thebibliography}{10}
\expandafter\ifx\csname url\endcsname\relax
  \def\url#1{\texttt{#1}}\fi
\expandafter\ifx\csname urlprefix\endcsname\relax\def\urlprefix{URL }\fi

\bibitem{FL}
{\rm M.A}.~Franey, {\rm W.G}.~Love, Phys.\ Rev.\ C 31 (1985) 488.

\bibitem{Doering}
R.~Doering, A.~Galonsky, D.~Patterson, G.~Bertsch, Phys.\ Rev.\ Lett. 35 (1975)
  1691.

\bibitem{ikeda}
K.~Ikeda, S.~Fujii, {\rm J.I}.~Fujita, Phys.\ Lett.\ 3 (1963) 271.

\bibitem{ikedasum}
C.~Gaarde{\it ~et al.}, Nucl.\ Phys. A334 (1980) 248.

\bibitem{Gaarde}
C.~Gaarde{\it ~et al.}, Nucl.\ Phys. A369 (1981) 258.

\bibitem{OsetRho}
E.~Oset, M.~Rho, Phys.\ Rev.\ Lett. 42 (1979) 47.

\bibitem{Ericson}
M.~Ericson{\it ~et al.}, Phys.\ Lett. 45B (1973) 19.

\bibitem{Shimizu}
M.~I. K.~Shimizu, A.~Arima, Nucl.\ Phys. A226 (1972) 282.

\bibitem{Bertsch}
{\rm G.F}.~Bertsch, I.~Hamamoto, Phys.\ Rev.\ C 26 (1982) 1323.

\bibitem{WA1}
T.~Wakasa{\it ~et al.}, Phys.\ Rev.\ C 55 (1997) 2909.

\bibitem{Shen}
S.~Quing-biao, F.~Da-chun, Z.~Yi-zhong, Phys.\ Rev.\ C 43 (1991) 2773.

\bibitem{Sasano}
M.~Sasano{\it ~et al.}, private communication.

\bibitem{Raywood}
{\rm K.J}.~Raywood{\it ~et al.}, Phys.\ Rev.\ C 41 (1990) 2836.

\bibitem{Yako}
K.~Yako{\it ~et al.}, Nucl.\ Phys. A684 (2001) 563c.

\bibitem{MatsuokaNoro}
N.~Matsuoka, T.~Noro, RCNP annual report 1987, p.176.

\bibitem{NoroOkihana}
A.~Okihana{\it ~et al.}, RCNP annual report 1987, p.171.

\bibitem{SAID}
{\rm R.A}.~Arndt, {\rm L.D}.~Roper, {\rm Scattering Analysis Interactive
  Dial-in (SAID) program, SM94 phase shift}.

\bibitem{dw81}
{\rm M.A}.~Schaeffer, J.~Raynal, {\rm Program DWBA70 (unpublished); J. R.
  Comfort, Extended version DW81}.

\bibitem{CooperHama}
{\rm E.D}.~Cooper{\it ~et al.}, Phys.\ Rev.\ C 47 (1993) 297.

\bibitem{WoodsSaxon}
A.~Bohr, {\rm B.R}.~Mottelson, Benjamin, New York, 1969, Nuclear structure,
  Vol.\ 1, p.239.

\bibitem{Kasagi}
J.~Kasagi{\it ~et al.}, Phys.\ Rev.\ C 28 (1983) 1065.

\bibitem{Horton}
{\rm J.L}.~Horton{\it ~et al.}, Nucl.\ Phys.\ A190 (1972) 362.

\bibitem{Preedom}
{\rm B.M}.~Preedom{\it ~et al.}, Phys.\ Rev. 166 (1968) 1156.

\bibitem{Graue}
A.~Graue{\it ~et al.}, Nucl.\ Phys. A187 (1972) 141.

\bibitem{TaddeucciDWIA}
{\rm T.N}.~Taddeucci{\it ~et al.}, Nucl.\ Phys. A469 (1987) 125.

\bibitem{bootstrap}
{\rm S.A}.~Teukolsky, {\rm W.T}.~Veterling, {\rm B.P}.~Flannery, ``Numerical
  Recipes in C++'', (Cambridge Univ., Cambridge, 2002) p.696.

\bibitem{Horowitz}
{\rm D.P}.~Murdock, {\rm C.J}.~Horowitz, Phys.\ Rev.\ C, {\bf 35}, 1442 (1987);
  C.J. Horowitz, D.P. Murdock, and B.D. Serot, in ``Computational Nuclear
  Physics 1'', edited by K. Langanke, J.A. Maruhn, and S.E. Koonin (Springer,
  New York, 1993) p.129.

\bibitem{RH}
L.~Rikus, N.~Nakano, H.~V. von Geramb, Nucl.\ Phys.\ {\bf A414}, 413 (1984);
  J.J. Kelly, Program Code {\sc LEA}.

\bibitem{Rijsdijk}
{\rm G.A}.~Rijsdijk{\it ~et al.}, Phys.\ Rev.\ C 48 (1993) 1752.

\bibitem{HamamotoSagawa}
I.~Hamamoto, H.~Sagawa, Phys.\ Rev.\ C 62 (2000) 024319.

\bibitem{Auerbach}
A.~Klein, {\rm W.G}.~Love, N.~Auerbach, Phys.\ Rev.\ C 31 (1985) 710.

\bibitem{Prout}
{\rm D.L}.~Prout{\it ~et al.}, Phys.\ Rev.\ C 63 (2000) 014603.

\bibitem{Conde}
H.~Cond\'{e}{\it ~et al.}, Nucl.\ Phys. A545 (1992) 785.

\bibitem{SuzukiSakai}
T.~Suzuki, H.~Sakai, Phys.\ Lett.\ B 455 (1999) 25.

\bibitem{ChewLow}
{\rm G.F}.~Chew, {\rm F.E}.~Low, Phys.\ Rev.\ 101 (1956) 1570.

\bibitem{Dickhoff}
{\rm W.H}.~Dickhoff{\it ~et al.}, Phys.\ Rev.\ C 23 (1981) 1154.

\bibitem{Arima2}
A.~Arima, W.~Bentz, T.~Suzuki, T.~Suzuki, Phys.\ Lett.\ B 499 (2001) 104.

\bibitem{OsetTokiWeise}
E.~Oset{\it ~et al.}, Phys.\ Rep. {\bf 83}, 281 (1982).

\bibitem{SagawaLeeOhta}
H.~Sagawa{\it ~et al.}, Phys.\ Rev.\ C 33 (1986) 629.

\bibitem{Nakano}
M.~Nakano{\it ~et al.}, J.\ Mod.\ Phys.\ E {\bf 10} (2001) 459; T. Tatsumi{\it
  ~et al.}, private communication.

\end{thebibliography}

\newpage

\end{document}